\documentclass[manuscript]{emulateapj}
\usepackage{natbib}

\newcommand{\lya}{Ly$\alpha$}
\newcommand{\ha}{H$\alpha$}
\newcommand{\hb}{H$\beta$}
\newcommand{\heii}{He~{\small II}}
\newcommand{\civ}{C~{\small IV}}

\newcommand{\ecsa}{erg cm$^{-2}$ s$^{-1}$ \AA$^{-1}$}
\newcommand{\ecs}{erg cm$^{-2}$ s$^{-1}$}
\newcommand{\es}{erg s$^{-1}$}

\newcommand{\kms}{km s$^{-1}$}

\slugcomment{Apj, accepted}

\shorttitle{A $z=2.38$ \lya\ nebula}  
\shortauthors{Scarlata et al.}

\begin{document}


\title{\heii\ emission in \lya\ nebulae: AGN or cooling
  radiation?\altaffilmark{1}}


\author{C. Scarlata, J. Colbert, H. I. Teplitz, C. Bridge}

\affil{{\it Spitzer} Science Center, 314-6, Pasadena, CA-91125}

\author{ P. Francis\altaffilmark{2}, P. Palunas\altaffilmark{3},
  B. Siana\altaffilmark{4}, G.~M. Williger\altaffilmark{5,6,7} and
  B. Woodgate\altaffilmark{8}}

\altaffiltext{1}{Based in part on observations obtained with the {\it
    Spitzer Space Telescope}, which is operated by JPL, California
  Institute of Technology for the National Aeronautics and Space
  Administration}

\altaffiltext{2}{The Australian National University}
\altaffiltext{3}{Carnegie Observatories}
\altaffiltext{4}{Caltech}
\altaffiltext{5}{University of Louisville}
\altaffiltext{6}{Johns Hopkins University}
\altaffiltext{7}{Catholic University}
\altaffiltext{8}{NASA/Godard Space Flight Center, USA}



\begin{abstract}
  We present a study of an extended \lya\ nebula located in a known
  overdensity at $z\sim 2.38$.  The data include multiwavelength
  photometry covering the rest-frame spectral range from 0.1 to
  250$\mu$m, and deep optical spectra of the sources associated with
  the extended emission.  Two galaxies are associated with the \lya\
  nebula. One of them is a dust enshrouded AGN, while the other is a
  powerful starburst, forming stars at $\sim 600 M_{\odot}$
  yr$^{-1}$. We detect the \heii\ emission line at 1640~\AA\ in the
  spectrum of the obscured AGN, but detect no emission from other
  highly ionized metals (\civ\ or N~{\small V}) as is expected from an
  AGN. One scenario that simultaneously reproduces the width of the
  detected emission lines, the lack of \civ\ emission, and the
  geometry of the emitting gas, is that the \heii\ and the \lya\
  emission are the result of cooling gas that is being accreted on the
  dark matter halo of the two galaxies, Ly1 and Ly2. Given the
  complexity of the environment associated with our \lya\ nebula it is
  possible that various mechanisms of excitation are at work
  simultaneously.

\end{abstract}




 
\section{Introduction}

Searches for high redshift \lya\ emitters are now successfully finding
large samples of normal star-forming galaxies at redshifts up to
$z\sim 6$.  These searches are also discovering rare, extremely
powerful and spatially large \lya\ sources, often referred to as
``blobs'' in the literature \citep{steidel2001}.  \lya\ nebulae are
characterized by extended \lya\ emission on scales up to hundreds of
kiloparsecs. The total integrated luminosity in nebulae can exceed
$10^{43}$ \es, that is, more than $10^{10}L_{\odot}$ of energy is
emitted in just the \lya\ line itself. These nebulae are similar both
in energy output and morphology to the emission line nebulae
associated with powerful high redshift radio galaxies, however no
radio sources have been detected in their proximity. The number
density of these sources is still poorly known, mainly because of the
rarity of these sources; \citet{yang2008} find only 4 \lya\ nebulae in
a survey targeting $z=2.3$ over an area of $\sim 5$ deg$^2$.  The
\lya\ nebulae were first discovered by \citet{steidel2001}, \citep[see
also ][]{francis1996,francis1997}, in a region characterized by an
overdensity of galaxies by about a factor of 6 when compared to the
field population.  Since then, most other nebulae have also been
discovered in overdense regions of the Universe
\citep{prescott2008,matsuda2004}.

Apart from their extended size, what makes \lya\ nebulae peculiar
compared to normal star forming \lya\ emitters is the lack of an
obvious source of ionization for the gas.  Various mechanisms have
been proposed to explain the observed emission, including
photoionization by either powerful AGNs or starbursts \citep[with the
ionizing radiation escaping toward the nebula, rather than the
observer;][]{steidel2001}, shocks produced by supernova driven winds
\citep{taniguchi2000,mori2004}, and cooling radiation from gas falling
into a dark matter halo \citep{fardal2001,yang2006}.

The latter model is of particular interest in light of recent
theoretical studies showing that a substantial fraction of the
in-falling gas does not shock heat to the virial temperature of the
dark matter halo, but arrives at the central galaxy predominantly
along filaments in the cosmic web
\citep{keres2005,dekel2006,dekel2008,brooks2008}. Such cold streams
could be responsible for the observed \lya\ nebulae
\citep{yang2006}. This mode of gas accretion is predicted to be the
way by which massive galaxies at $z\sim 2$ can still acquire gas that
can be turned into stars. However, an observational confirmation of
this mechanism is still missing, although a few possible examples have
been presented in the past few years
\citep{nilsson2006,smith2007,smith2008}.

Here we present an in--depth study of a complex \lya\ nebula at $z
\sim 2.38$ found to be associated with multiple 24$\mu$m sources
\citep{palunas2004,colbert2006}. This \lya\ nebula (RA$=21:42:42.72$,
DEC$=-44:30:00.0$) is located within a large structure of \lya\
emitting galaxies, in an $80\times 80\times 60$ comoving Mpc region
surrounding the known $z=2.38$ group of galaxies J2143--4423
\citep{francis1996}.  The structure is defined by $\sim$ 30
spectroscopically confirmed \lya\ emitters \citep{francis2004}, a
space density a factor of $5.8\pm 2.5$ greater than that found by
field samples at similar redshifts.  The distribution of the \lya\
emitters is possibly filamentary, and contains several $5-10$ Mpc
scale voids. This group of galaxies was first discovered as a cluster
of Lyman limit absorption-line systems along the line of site of two
$z>3$ QSOs \citep[][]{francis2001,francis1993}.

The paper is organized as follows. Section~\ref{sec:data} describes
the imaging and spectroscopic data, Section~\ref{sec:results} present
the results. The origin of the galaxies associated with the nebula is
studied in Section~\ref{sec:sed}, while in
Section~\ref{sec:discussion} we discuss various plausible mechanisms
for the gas ionization. Section~\ref{sec:conclusions} presents our
conclusions. Throughout the paper we assume $\Omega_m = 0.3$,
$\Omega_m + \Omega_{\Lambda} = 1$, and $H_0 = 70$ km s$^{-1}$
Mpc$^{-1}$. All magnitudes are AB magnitudes \citep{oke1974}, unless
otherwise specified.

\section{Observations and data analysis}
\label{sec:data}

\subsection{Imaging data}
The J2143--4423 field is a well studied area of the sky, with deep
imaging from X-ray to the radio wavelengths (Table~\ref{tab:data}).
This wealth of data allows an in depth study of the \lya\ nebula.  The
optical and NIR data, including $U,\, B,\, G,\, R$, narrow band
centered at 4107\AA, $J$ and $H$ images were acquired at the Cerro
Tololo Interamerican Observatory (CTIO) 4m telescope, using the
Mosaic~II imager, and the near Infrared Side Port Imager camera
(ISPI). Mid-IR images of the field were taken with the four channels
of the Infrared Array Camera \citep[IRAC,][]{fazio2004} at 3.6, 4.5,
5.8 and 8.0$\mu$m and with the Multiband Imaging Photometer
\citep[MIPS,][]{rieke2004} at 24$\mu$m on the {\it Spitzer}
Telescope. The data reduction of the optical and infrared data will be
discussed in detail in a companion paper presenting the stellar
population properties of all \lya\ emitters identified in the cluster
field \citep[][in preparation]{scarlata2009}. Observations at 850
$\mu$m were obtained with CCAT$+$LaBoca installed on the Atacama
Pathfinder EXperiment and presented in \citet{beelen2008}. X-ray data
covering 0.5-10 keV were acquired with Chandra+ACIS \citep[][in
preparation]{williger2008}. Radio observations at 1334 MHz were
acquired with the compact array of the Australian telescope
\citep{francis1996}. No radio sources are detected (apart from the
$z=3$ QSO 2139-4434) in the field, down to a 3$\sigma$ limit of 3.3
mJy.

\begin{figure}
\epsscale{1.20}
\plotone{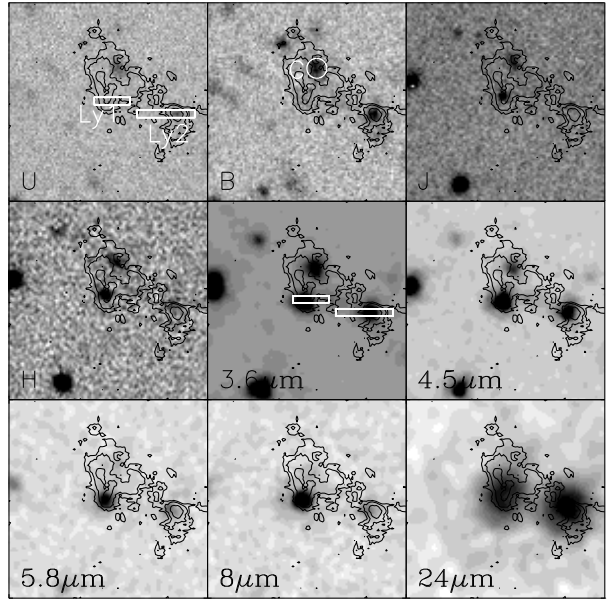}[t]
\caption{\label{fig:images} Images of the area around the \lya\
  nebula. The images are 27 arcseconds on the side (corresponding to
  $\sim 220$ kpc at $z=2.37$). From top left, to bottom right, we show
  $U,\,B,\,J,\,H,$ IRAC channels $1,\,2,\,3,\,4$ and MIPS 24$\mu$m.
  The contours in each image show the diffuse \lya\ emission, and
  represent $3,\,4$, and $5\times \sigma_{Ly\alpha}$
  ($\sigma_{Ly\alpha}=2\times 10^{-18}$ \ecs\ arcsec$^{-2}$). The
  white boxes overplotted to the $U$ and 3.6$\mu$m images show the
  positions and orientation of the slits. The names of the two $z=2.38$
  galaxies are indicated near the slit. The position of the foreground
  object, $C$, is also shown.}
\end{figure}

In Figure~\ref{fig:images} we show cut outs of the multiband optical
through MIR images, centered at the position of the \lya\
nebula\footnote{For displaying purposes only, we show all images with
  the same orientation and pixel scale. The photometry measurements
  were performed on the original exposures.}. The stamps are 220 kpc
on a side at the redshift of the nebula. The \lya\ emission, indicated
with the solid contours in Figure~\ref{fig:images}, covers three
continuum sources indicated in the figure as Ly1, Ly2 and C.
Multislit spectroscopy obtained with the Inamori Magellan Areal Camera
and Spectrograph (IMACS) at the Magellan Baade telescope in August
2008 showed that source C is a foreground galaxy at redshift $z=0.82$.
Hereafter we will only consider the two sources Ly1 and Ly2, that are
found to be at the same redshift of the extended \lya\ nebula (see
below).

The available data allow a detailed description of the spectral energy
distribution (SED) of the sources detected within and in the vicinity
of the $z=2.38$ \lya\ nebula. Particular care needs to be taken in
measuring the multiwalength SED of the objects due to the
intrinsically different spatial resolution of the various
datasets. Since both Ly1 and Ly2 appear unresolved even in the best
quality images available (0\farcs9 full width half maximum, FWHM, in
the J and H images), we computed their total flux within a circular
aperture with a diameter optimized independently for each band. The
aperture size was chosen to be large enough to include as much flux as
possible from the source, while maximizing the signal-to-noise ratio
within the aperture and minimizing the contamination by nearby
sources. The aperture diameter was typically chosen to be $1.5\times$
the FWHM. The aperture magnitudes were then transformed into total
magnitudes by applying an aperture correction derived for each image
using isolated pointlike sources. We also computed Kron-like
elliptical aperture total magnitudes using SeXtractor
\citep{sex}, and we verified that the results do not differ
from those obtained with circular apertures. The broad band total
magnitudes for Ly1 and Ly2 are shown in Table~\ref{tab:photometry}.

\subsection{Spectroscopic data}
Multi object spectroscopy was performed in the region surrounding the
\lya\ nebula using the Gemini Multi Object Spectrograph
\citep[GMOS][]{hook2003} on Gemini South.  The optical spectroscopy
targeted the objects associated with the nebulae, and nearby 24$\mu$m
sources. The spectra were obtained with the B600 grating, tilted at
two different angles (corresponding to central wavelengths of 4820 and
4870\AA). With this setup we were able to cover the entire spectral
range between 3700 and 6000~\AA, removing the gap between the two CCDs
of the camera. In order to increase the signal to noise ratio of the
spectra, we binned the images by $2 \times 2$ pixels during the read
out.  The pixel scale of the CCD is 0\farcs0727 pixel$^{-1}$ and the
seeing conditions were 0\farcs8.  Thus, the binning does not cause any
significant loss in either spatial or wavelength resolution.  After
binning they are 0\farcs15 pixel$^{-1}$ and 0.9 \AA\ pixel$^{-1}$,
respectively.  The mask was observed for a total of 4.7 hours
($8\times 2100s$), with clear conditions and seeing of 0\farcs8 FWHM.
The slitlets in the mask were 1\farcs0 wide.

The data were reduced using the {\tt gsmos} task included in the {\tt
  IRAF} gemini package. Briefly, the reduction steps include bias
subtraction, slit by slit flat-field normalization, wavelength
calibration, and 2d spectral rectification. Wavelengths were
calibrated using the arc line exposures, yielding a solution with an
rms error of approximately 0.3\AA. The sky was measured in each slit
by fitting a function (either a constant or an order 2 Chebyshev
function) to a selected sample of pixels in each column of the
slit. One dimensional spectra were extracted with {\tt gsextract} in
{\tt IRAF}\footnote{IRAF is distributed by the National Optical
  Astronomy Observatories, which are operated by the Association of
  Universities for Research in Astronomy, Inc., under cooperative
  agreement with the National Science Foundation.} \citep{tody1993}.

\begin{figure}[!t]
  \epsscale{1.}  \plotone{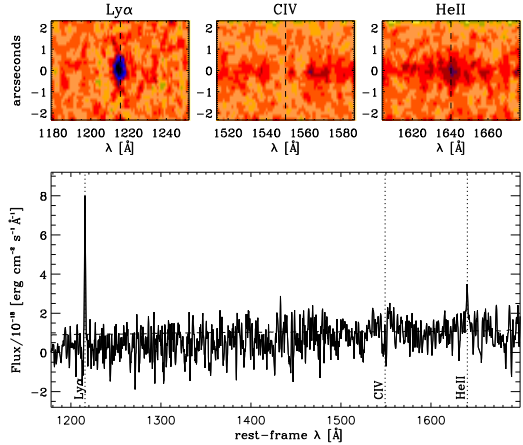} \caption{\label{fig:specA} GMOS
    rest frame UV spectrum of Ly1. The extracted spectrum is shown in
    the bottom panel (solid line), and the 1 $\sigma$ error is shown
    with a dashed line. The spectrum was smoothed with a boxcar
    average over 2.7\AA. The dotted vertical lines show the position
    of the \lya, \civ, and \heii\ emission lines. In the upper part of
    the Figure, we show portions of the 2 dimensional spectra centered
    at the wavelength of the observed emission line features, namely
    \lya\ and \heii, and at the expected wavelength of the \civ\
    line.}
\end{figure}

The spectral resolution, measured from the emission lines of the lamp
spectra, is 5.0 \AA\ (FWHM), corresponding to a velocity resolution of
$\sim 310$ \kms\ at 4850\AA. We removed the wavelength-dependent
instrumental response using the response curve derived from the
spectrum of a standard star. The spectra were flux calibrated using
the broadband $B$ magnitudes of Ly2 (see Figure~\ref{fig:images}). We
convolved the extracted spectrum with the $B-$band filter transmission
curve to obtain the measured flux density ($f_{measured}$). The ratio
$\epsilon$ between $f_{measured}$ and the known $B-$band flux density
provides the flux calibration for the spectra ($f_{\lambda ,
  B}=f_{measured}/\epsilon$).  Such a calibration also includes the
correction for any flux lost because of the finite size of the
slit. Due to the proximity of Ly1 and Ly2, we applied the same
calibration factor to both sources. From the error on the broad band
magnitude, we estimate a $15$\% error on the flux calibration.

We obtained mid-infrared spectra for both Ly1 and Ly2 with the {\it
  Spitzer} Infrared Spectrograph (IRS), using the 1st order of the
Long-Low module (LL1), which is sensitive from $19.5-38.0$~$\mu$m. The
LL1 module has a spatial resolution of 5\farcs1 pixel$^{-1}$ and a
wavelength resolution of $R=58-116$. The IRS data were acquired on
June 13, 2007 (GO-30600, P.I. J. Colbert), using the IRS Spectral
Mapping Astronomical Observation Template, placing each source at six
separate positions along the length of the slit, separated by 20$''$.
We used 10 mapping cycles and the 120 second ramp exposure for both,
producing a total integration time of 120 minutes for each
source. After sky subtraction and correction for latent charge
build-up, the data at each position were averaged together and
extracted using the {\it Spitzer} Science Center program SPICE
(v.2.1.2) and its optimal extraction option. We then averaged all six
spectrum positions together before making a final calibration
adjustment based on the known 24~$\mu$m fluxes. For further data
reduction details see \citet[][in preparation]{colbert2009}.

\begin{figure}[!t]
\epsscale{1.}
\plotone{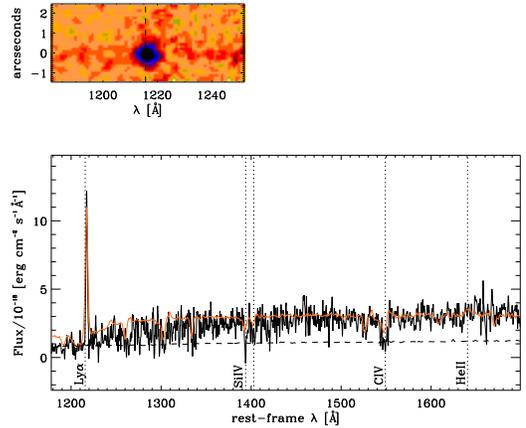} \caption{\label{fig:specB}
  Same as Figure~\ref{fig:specB} for Ly2. The orange spectrum is the
  composite spectrum derived for $z\sim 3$ LBG by
  \citet{shapley2003}.}
\end{figure}

\section{Results}
\label{sec:results}
In Figure~\ref{fig:images}, we can see that the \lya\ emission extends
over an area of about $\sim 100$ square arcseconds.  It encompasses
the two sources of continuum, indicated as Ly1 and Ly2, and shows
peaks centered on them.  Although the bulk of the extended \lya\ 
emission is localized nearby Ly1, this galaxy is not located at the
center of the emission. The total \lya\ flux integrated over the
entire nebula is $1.5\times 10^{-15}$ \ecs\ \citep{palunas2004},
corresponding to a total luminosity $L_{Ly\alpha}=6.5\times 10^{43}$
erg s$^{-1}$ ($1.6\times 10^{10} L_{\odot}$).

The GMOS slits were positioned over the continuum sources, as
indicated by the white boxes in Figure~\ref{fig:images}. The extracted
spectra of Ly1 and Ly2 are shown in Figure~\ref{fig:specA}
and~\ref{fig:specB}, respectively. In the bottom panel of each figure
we show the extracted spectrum with the 1 $\sigma$ associated error.
In the upper part of the Figure, we show portions of the two
dimensional spectra centered at the wavelength of the main features
considered in this work, namely \lya, \civ, and \heii\ $\lambda~1640$
for Ly1 and \lya\ for Ly2. The spectra confirm that both sources are
at redshift $z\sim 2.37$, and not a chance superposition with the
\lya\ nebula.
\begin{figure}[t]
  \epsscale{1.}  \plotone{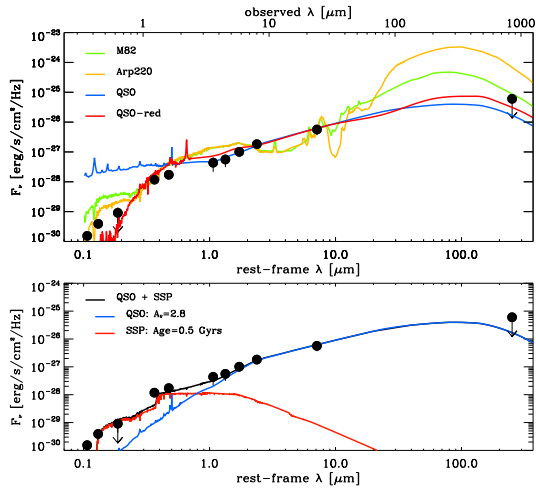} \caption{\label{fig:sedA} Top
    panel: Spectral energy distribution of Ly1 (circles), compared
    with template SEDs of different (both active and star-forming)
    galaxies from the compilation of templates by
    \citet{polletta2006}. All templates are normalized to the observed
    24$\mu$m flux. Bottom panel: comparison of Ly1 SED (circles) with
    a composite model (black line) derived by combining a QSO template
    with an $A_V=2.8$ magnitudes (blue curve) plus a 0.5 Gyr old
    stellar population (red curve).}
\end{figure}

Apart from the narrow \lya\ emission line, the spectrum of Ly1 shows
an emission line at $\lambda=5531.7$\AA, corresponding to \heii\ at
redshift $z=2.373$. No \civ\ $\lambda~1550$ emission line is detected
in the spectrum. To derive an upper limit to the intensity of the
\civ\ emission we added progressively fainter emission lines at the
expected wavelength of the line, until it could no longer be
identified at the $2\sigma$ level. As can be seen in the two
dimensional spectrum of Ly1, neither the \lya\ nor the \heii\ emission
lines show an extended morphology, and are cospatial with the
continuum.  The slit, however, was oriented perpendicular to the bulk
of the extended emission, so that it covered only the most southern
clump (see Figure~\ref{fig:images}). Not knowing the morphological
extension of the \heii\ emission, the total luminosity derived from
the observed spectrum can be regarded as a lower limit, since the
\heii\ emission could be extended. Estimation of the size of the
\heii\ emitting region is limited by the seeing (FWHM$=$0\farcs8)
during the observations. The size of the nebula is spatially
unresolved so its size is $<\sim 6.5$ kpc.

Ly2 has a bright rest--frame UV continuum, with a $B-$band
(corresponding to $\sim$1300\AA\ rest-frame) AB magnitude of 23.8, and
we are able to identify various absorption lines in its spectrum. The
features that we detect come from both low- and high ionization
interstellar lines (e.g., SI~{\small II~1526},C~{\small II~1334},
Si~{\small IV~1393-1402}, C~{\small IV~1550}), typical of high
redshift star-forming galaxies \citep{shapley2003}. The redshift of
Ly2 measured using the absorption lines is $z=2.362$, slightly lower
than the redshift estimated from the \lya\ emission line
($z_{Ly\alpha}=2.367$). Such a difference corresponds to a velocity
shift of $\sim$ 450 km s$^{-1}$, and is in the range of the typical
shifts observed in LBGs \citep[$\langle \Delta v\rangle=650$ \kms,
][]{shapley2003}. The exact difference, however, is uncertain due to
the absorption by the \lya\ forest on the blue side of the line.

The redshifts of the two galaxies place them within the structure of
\lya\ emitters.  Spectroscopy confirmed the structure to be at a mean
redshift of $z=2.378$, with a standard deviation of 820 \kms\
\citep{francis2004}.

The observed equivalent widths (EWs), velocity dispersion, and line
fluxes of the emission lines detected in Ly1 and Ly2 are presented in
Table~\ref{tab:spectroscopy}.  The luminosity of the \lya\ emission
measured from both Ly1 and Ly2 accounts for $\sim 10$\% of the total
\lya\ luminosity integrated over the extended nebula. We computed the
mean flux density at rest--frame 1500\AA\ for both Ly1 and Ly2 by
averaging the observed flux within a band of $\pm 20$\AA.  Ly1 and Ly2
have $f_{1500}=8 \pm 5 \times 10^{-19}$ and $2.8\pm 0.8 \times
10^{-18}$ \ecsa, respectively.

\section{The origin of the two continuum sources}
\label{sec:sed}
\begin{figure}[!t]
\epsscale{1.}
\plotone{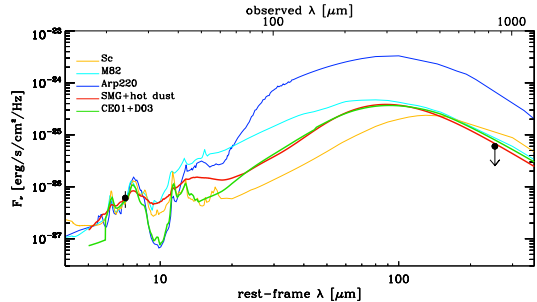} \caption{\label{fig:sedB}
  Comparison between the IR photometric points of Ly2 at observed
  24$\mu$m and 850$\mu$m with various starburst templates taken from
  \citet{polletta2007} and \citet{pope2008}.}
\end{figure}

The SEDs of the two objects associated with the \lya\ nebula can
provide valuable information on the mechanisms responsible for the
ionization of the gas.  Both sources are characterized by high mid--IR
fluxes ($F_{24 \mu m}\sim 0.6$mJy) and are not detected at the submm
wavelength. The complete SEDs of Ly1 and Ly2 are shown in
Figures~\ref{fig:sedA} and \ref{fig:sedB}. The LaBoca 3$\sigma$ flux
limit of 6 mJy at $850\mu$m is also shown.

{\bf Ly1:} Ly1 is a source of intense emission at 24 $\mu$m,
corresponding to a $\nu L_{\nu}\sim 2.3\times 10^{11} L_{\odot}$ at
rest-frame $\sim 7 \mu$m.  The strong mid-infrared flux and its
proximity (9\farcs5, corresponding to $\sim 80$ kpc at $z=2.38$) to
the other powerful 24$\mu$m source, Ly2, have been suggested to imply
that both these galaxies are undergoing a strong burst of star
formation, possibly induced by their interaction \citep[e.g.,
][]{colbert2006}.  Other evidence, however, indicates that Ly1 is a
heavily reddened QSO. First, the SED of Ly1 is characterized by a
power law shape from 1 to 7$\mu$m rest--frame, and by a high
mid-infrared to optical flux ratio ($f_{24}/f_R \sim 10^3$). These
properties, together with the powerful 24$\mu$m emission of this
object, are typical of obscured AGNs as recently suggested by
\citet{fiore2008}. In Figure~\ref{fig:sedA}, top panel, we compare the
SED of Ly1 with templates of active galaxies \citep[taken
from][]{polletta2007} and local starburst galaxies. All templates are
normalized to the 24$\mu$m observed flux density.  Among the Polletta
et al. templates, we show the templates of a reddened and of an
un-obscured QSO (red and blue curve, respectively). We do not show the
templates of the low luminosity Seyfert~2 galaxies, since their SEDs
show a pronounced minimum at $\sim 3\mu$m that is not consistent with
the data of Ly1. We also show Arp220 (yellow curve) and M82 (green
curve), since these galaxies have, respectively, the reddest and the
bluest $f_{250\mu m}/f_{7\mu m}$ rest-frame color.

The shape of the observed SED of Ly1 is clearly inconsistent with
Arp220 and M82. At wavelengths shorter than $10\mu$m rest-frame, both
the M82 and the Arp220 templates are characterized by the 1.6$\mu$m
bump in the SED due to the minimum in the H$^-$ opacity, and by a
minimum in the flux density around $3\mu$m. The minimum at $3\mu$m
marks the transition between stellar- and dust-dominated emission. The
fact that it is not observed in the Ly1 SED is a strong indication of
the presence of a hot dust component, whose emission dominates in the
3 to 5$\mu$m rest frame, and is usually absent in star-formation
dominated galaxies.

The optically selected un-reddened QSO template reproduces well the
SED of Ly1 at $\lambda > 1\mu$m, however it strongly over-predicts its
UV energy output. A better agreement to the observed data at short
wavelengths is given by the SED of the reddened QSO model, but the UV
through optical rest--frame color of this template is too red to match
the observed data. The lack of broad emission lines in the optical
spectrum of Ly1 indicates that the QSO is heavily obscured, making it
likely that the continuum in the optical/UV rest-frame is dominated by
the light of the host galaxy \citep[e.g.,][]{urrutia2008}. This is
shown in the bottom panel of Figure~\ref{fig:sedA}, where we model the
SED of Ly1 using a combination of a reddened QSO (blue curve) plus a
0.5 Gyr old stellar population template (red curve). To reproduce the
observed SED, the QSO template needs to be reddened by an extinction
of $A_V=2.8$ magnitudes. Assuming the Galactic dust--to--gas ratio
\citep[$E(B-V)/N_H=1.7\times 10^{-22}$ cm$^{-2}$;][]{maiolino2001}, the
column density of Hydrogen would be $\sim 4\times 10^{21}$
cm$^{-2}$. This estimate represents a lower limit, since the
dust--to--gas ratio in AGNs is nearly always found to be a factor
between 3 and 100 lower than the Galactic one \citep[corresponding to
$N_H$ between $1.2\times 10^{22}$ and $4\times 10^{23}$
cm$^{-2}$][]{maiolino2001}.

Ly1 is not detected in the Chandra images down to a 2$\sigma$ limit of
$f_{0.5-7keV}= 10^{-15}$ \ecs\ \citep[corresponding to $L(2-10KeV)\sim
3\times 10^{43}$ erg s$^{-1}$, assuming a power law spectrum with
energy index $\alpha_E=0.8$, ][in preparation]{williger2008}.  If Ly1
was unabsorbed, its 6 $\mu$m luminosity would imply a rest-frame 2--10
KeV luminosity of a few 10$^{44}$ erg s$^{-1}$
\citep{fiore2008,lutz2004}, i.e., above the observed Chandra flux
limit.  To absorb the X-ray luminosity to the observed Chandra limit
requires a gas column density  of $N_H$ of $\sim 5\times 10^{22}$
cm$^{-2}$, consistent with the range of $N_H$ derived in the previous
paragraph.

The {\it Spitzer} IRS spectrum of Ly1 reveals that the integrated
contribution of the polycyclic aromatic hydrocarbons (PAH) emission
lines to the total mid-IR flux between 7 and 9$\mu$m rest--frame is
less than 10\% \citep[][in preparation]{colbert2009}. Strong PAH
emission is usually observed in strongly star forming objects, and
dominates the MIR spectrum of typical local starbursts.

Although the observed SED of Ly1 is consistent at all wavelengths with
that of a reddened QSO, the optical spectrum of Ly1 (see
Figure~\ref{fig:specA}) is difficult to explain as a result of
photoionization by the QSO continuum. In particular, although the
presence of the \heii\ emission is typical of ``type 2'' QSO spectra,
the absence of the \civ\ emission is puzzling. We will return to this
point in Section~\ref{sec:discussion}.

\begin{figure}
\epsscale{1.}
\plotone{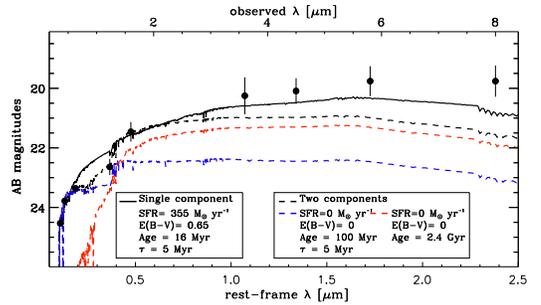} \caption{\label{fig:sedBfitting}
  Results of the SED fitting to the rest-fame UV to optical photometry
  of Ly2. The black solid line shows the result of the one component
  fit, while the dashed line represents the combined spectrum of an
  old stellar population (red dashed curve) and a young starforming
  population (blue dashed curve). The parameters of the two models are
  indicated in the inset. }
\end{figure}

{\bf Ly2:} The rest-frame UV and mid-IR SEDs of Ly2 are shown in
Figures~\ref{fig:sedBfitting} and \ref{fig:sedB}. Although extremely
powerful at 24~$\mu$m (Ly2 has the same luminosity as Ly1 at
24~$\mu$m), the majority of the energy in the mid-IR is due to
young-star radiation reprocessed by dust.  This is confirmed by both
the shape of the rest-frame IR SED, and by the mid-IR spectrum of Ly2,
shown in Figure~\ref{fig:PAH}. The IR SED shows the H$^-$ opacity
minimum at 1.6~$\mu$m, and the transition to dust dominated emission
at about 3~$\mu$m. Indeed, the observed $f_{8.0}/f_{4.5}$ color is
less than 2, which is typical of $z\sim 2$ star-formation dominated
objects \citep{pope2008}.  The IRS spectrum shows prominent PAH
emission lines, whose integrated energy amounts to more than 60\% of
the mid-IR flux \citep[][in preparation]{colbert2009}, confirming the
presence of a strong burst of star-formation.  In the rest-frame UV,
the slope of the continuum measured from the $B-R$ color (i.e., $\sim
f_{1900}/f_{1300}$ rest-frame) is rather flat, with $B-R= 0.4\pm 0.2$
magnitudes. This indicates that the UV is dominated by a young stellar
population, with some amount of extinction \citep[][see
Section~\ref{sec:sfrLy2}]{meurer1999}. The observed $J$ and $H$ bands
bracket the 4000 \AA\ break at the redshift of Ly1. The $J-H$ of 1.1
magnitudes is however too red to be generated by a spectral break
alone, and requires substantial dust (see \ref{sec:sfrLy2}).

In Figure~\ref{fig:sedB} we compare the mid-IR photometry of Ly2 with
three different templates of starburst and normal star-forming
galaxies, namely Arp220 (blue curve), M82 (light blue curve), and a
late type spiral galaxy galaxy (Sc type, yellow curve). As in
Figure~\ref{fig:sedA}, all templates are normalized to the observed
24~$\mu$m flux density. The 24~$\mu$m flux corresponds to an energy
output at 7~$\mu$m rest-frame of $\nu L_{\nu}> 10^{11} L_{\odot}$,
i.e., for any assumed template SED, Ly2 would be classified as an
Ultra Luminous Infrared Galaxy (ULIRG, defined as a galaxy with
$L_{IR}>10^{12} L_{\odot}$). However, the typical local template for a
ULIRG, Arp220, clearly overestimates the flux at 850~$\mu$m. The
far-to mid-IR flux ratio of Ly2 is more consistent with that of local
starburst (such as M82) and normal galaxies (Sc).

In Figure~\ref{fig:sedB} we also show two templates derived by
\citep{pope2008} to reproduce the mid-to-far IR SED of dust obscured
galaxies at $z\sim 2$ \citep{desai2008}. These galaxies have
typical ULIRG luminosities, however their SED differs from the shape
of the local ULIRG template \citep[see
also][]{papovich2007,murphy2008}. The total IR luminosity derived
using the \citep{pope2008} templates is $L_{IR}= 7\times 10^{12}
L_{\odot}$. We will use the total IR luminosity in next section to
compute the star formation rate (SFR) in Ly2.

\subsection{SFR estimate in Ly2}
\label{sec:sfrLy2}
The energetic output of Ly2 is not dominated by emission coming from
an active nucleus, but rather from star-formation. The SFR in Ly2 can
be measured, or at least constrained, with indicators at different
wavelengths. Each of these indicators reflects a different emission
mechanism, e.g., direct young star light at 1500\AA\ rest frame, or
cold dust emission at far-IR wavelengths \citep{kennicutt1998}.

First, we derive the SFR by using the total IR luminosity derived
using those templates that satisfy the sub-mm flux limit discussed
above. We find consistent values for the estimated IR luminosity of $
\sim 3-6\times 10^{12} L_{\odot}$, using both the Sbc and the
\citet{pope2008} templates. Assuming the \citet{kennicutt1998}
conversion between the IR luminosity and the SFR, we find that Ly2 is
forming stars at a rate of 600-1200 $M_{\odot}$ yr$^{-1}$.

\begin{figure}[!ht]
\epsscale{1.}
\plotone{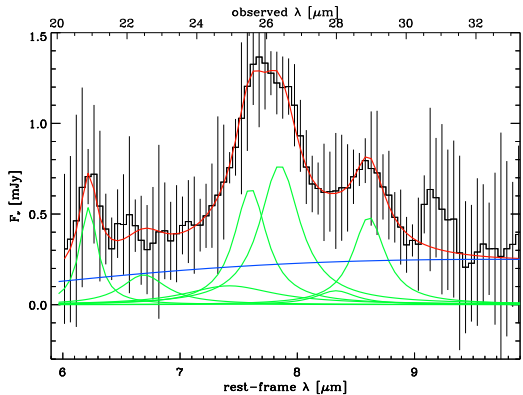} \caption{\label{fig:PAH}
  IRS spectrum of Ly2 shown as a function of rest-frame wavelength
  (black curve). The red curve shows the best fit model to the
  observed spectrum derived combining the PAH emission lines in the
  rest-frame 6 to 10$\mu$m (green curves) plus one power law continuum
  (blue curve). }
\end{figure}

A second way to estimate the IR luminosity is to use the integrated
intensity of the PAH emission lines \citep[e.g.][]{pope2008pah}.  The
IRS spectrum of Ly2 covers the rest frame wavelength range between 6
and 12~$\mu$m, and the clearest feature detected is the 7.7~$\mu$m
complex. In order to measure the individual PAH emission lines, we
used the {\tt PAHfit} software package \citep{smith2008}. Given the
limited coverage of the data, we only include in the fit the lines and
one red continuum. The fit to the Ly2 IRS spectrum is shown in
Figure~\ref{fig:PAH}.  We find an integrated luminosity in the
7.7~$\mu$m complex of $\sim 6.25\times 10^{10} L_{\odot}$. Using the
\citet[][]{pope2008pah} relation between the line luminosity and
$L_{IR}$ and the Kennicutt conversion to SFR, we find an SFR of $\sim
1500$ $M_{\odot}$ yr$^{-1}$, consistent with the value inferred from
the SED templates.  However, this estimate is subject to a large
uncertainty. Indeed, \citet{murphy2008} reanalyzed the
\citet[][]{pope2008pah} galaxy sample using a different fitting
method, and found $L_{IR}$ a factor of up to $\sim 4.8$
smaller. 

As a third approach to measure the SFR, we perform a complete SED fit
to the observed photometric points, including the data between the
rest-frame 0.1 to 3$\mu$m.  (The advantage of fitting the full SED
rather than using just the UV luminosity is that the former provides
the freedom to vary the star formation history and the age of the
stellar population). In the fits, we use the Starburst99 stellar
population models \citep{leitherer1999,vazquez2005}.  We consider a
Salpeter Initial Mass Function, and three star formation histories:
single burst stellar population, continuous star formation rate, and
exponentially declining star formation rate, with two values for the
e-folding time $\tau=150$ and $\tau=5$ Myr.  We let the age of the
stellar population vary, considering 60 steps between 1 Myr and 2.5
Gyr, evenly spaced in log space. We consider only models with a
metallicity of 0.4$Z_{\odot}$, consistent with metallicity
measurements of LBGs at similar redshift \citep{giavalisco2002}.  We
account for dust extinction using the dust extinction law of
\citet{calzetti1997}. We find that the best fit model is provided by a
young stellar population with an exponential star formation history
($\tau$ of 5 Myr, an age of $16$Myr, and $E(B-V)=0.65$). The
instantaneous SFR is 355~$M_{\odot}$ yr$^{-1}$. This model requires a
substantial amount of dust to reproduce the red rest-frame UV to
optical color.

Alternatively, the red colors of the SED of Ly2 could be due to the
presence of an old stellar population dominating the stellar mass. To
check for this, we also perform a two-component fit. The galaxy SED is
modeled by the sum of an old stellar population (parametrized with a
single burst SED and age of 2.5 Gyr\footnote{We note that a difference
  of $\pm$1 Gyr in the age of the old stellar population does not
  change the results.}), and a more recently star-forming population
(parametrized with an exponentially declining SFH with $\tau=5$
Myr). We let the age of the star-forming template vary, and we apply
dust extinction only to the star-forming component. In
Figure~\ref{fig:sedBfitting}, we show the best fit composite SED
together with the SEDs of the young and old population,
respectively. The age of the young component in the best--fit
composite model is of $1.4\times10^8$ Myr, resulting in a negligible
instantaneous SFR. This model can therefore be excluded, since the
24$\mu$m luminosity and the observed PAH emission clearly indicate
that Ly2 is actively forming stars.

To conclude, the analysis of the multi-wavelength data of the two
continuum sources associated with the \lya\ nebula shows that Ly1
hosts a powerful active nucleus heavily obscured in the rest-frame
optical and UV, and whose emission dominates for wavelength $>1 $
$\mu$m. Unlike Ly1, the energy output of Ly2 is dominated at all
wavelengths by the star-burst. The strength of the burst, quantified
with different SFR indicators, ranges from $\sim 400$ to $\sim 1500$
~$M_{\odot}$ yr$^{-1}$.

\section{Discussion}
\label{sec:discussion}

The source of energy capable of ionizing the gas in extended
\lya\ nebulae is still not understood. One of the main reasons for
this is that \lya\ is a resonant line with a very large cross section,
so that various effects, including the geometry of the ISM, the gas
velocity field, and the dust absorption, significantly alter the
observed \lya\ emission, both in its energy output and in the spectral
shape of the line. In order to properly model the observed emission, a
complete treatment of the radiation transfer of the \lya\ photons is
needed and this can be done only with simplified assumptions about the
dynamical state and chemical composition of the ISM. For these reasons
the \lya\ emission alone is often not enough to constrain the various
mechanisms that have been proposed to explain the source of energy in
the \lya\ nebulae.

The nebula presented here differs from most other known nebulae, due to
the presence of the \heii\ emission line in the spectrum.  Unlike the
\lya, \heii\ $\lambda 1640$ is not a resonant line and its output is
much less affected by the difficulties mentioned above, so that if
\heii\ is observed in conjunction with the \lya, its line width,
spatial distribution, and intensity can help establish the source of
energy in the emitting gas. The presence of the \heii\ line in the
spectrum is a strong constraint for the ionizing source since it
requires photons with enough energy to ionize He$^+$ (i.e., photons
with wavelength $\lambda < 228 $\AA, 54.4 eV).

In this section we will discuss possible mechanisms that can explain
the observed emission line spectrum. We will show that the \heii\
emission line is most likely originating from ionized gas located {\it
  outside} the Ly1 galaxy, along the line of sight to the observer.
Using the limit on the ratio between the \civ\ and \heii\ emission
line intensities as a diagnostic, we also exclude the possibility that
the gas is ionized by the continuum of the obscured AGN. Furthermore,
we will demonstrate that the possibility that the gas is ionized by a
starburst of population~III (or very low metallicity) stars is highly
unlikely. While not yet proven, the mechanism that simultaneously
explains all the observables is cooling radiation from gas falling on
the potential well of the dark matter halo of Ly1 and Ly2.

\subsection{\heii: stellar origin?}

As discussed in Section~\ref{sec:results}, we do not know whether the
\heii\ emission is extended (as the \lya\ is) or whether it is
concentrated on the Ly1 source. We need to consider the possibility
that it is of stellar origin, i.e., originating in hot and dense
stellar winds of Wolf-Rayet stars. The main observation that argues
against this possibility is the width of the \heii\ emission line
($\sigma < 172$ \kms). If the line were created in hot winds its width
would be much broader than what is measured in Ly1. In fact, typical
WR galaxies have \heii\ emission line widths on the order of thousands
of \kms\ \citep{schaerer2003}, while the uncorrected velocity
dispersion measured for the Ly1 \heii\ line is $\sigma <
200$~\kms. Furthermore, \citet{leitherer1995} showed that the emission
of \heii\ from hot winds is always associated with a strong P-Cygni
emission from \civ~{\small $\lambda 1550$}, always stronger than
\heii\ line emission.  As we said in Section~\ref{sec:results} we see
no \civ\ emission in the spectrum of Ly1.

\subsection{\heii: other mechanisms?}
The narrow velocity width of the \heii\ line suggests that it is more
likely to be of nebular origin. This further suggests that both \heii\
and \lya\ originate in gas ionized  by a hard continuum source. Only
two kinds of sources produce a UV continuum capable of ionizing
He$^{+}$: QSOs and extremely metal poor population~III stars.  As we
discussed in Section~\ref{sec:sed}, the two sources of continuum
radiation have broad band SEDs consistent with a reddened QSO (Ly1)
and a normal star forming galaxy (Ly2). Both are powerful 24$\mu$m
sources and emit a high fraction of their total energy in the mid- and
far-IR. 

\subsubsection{Evidence against an AGN}
The continuum of an active nucleus would be hard enough to photoionize
He$^+$ and produce the observed \heii\ line. We have shown in
Section~\ref{sec:sed} that Ly1 hosts a heavily obscured active
nucleus, so the possibility exists that the emission line spectrum is
produced by AGN photoionization. In this section we discuss various
pieces of evidence that suggest that the \heii\ and the \lya\ emission
lines are not the result of photoionization by a QSO continuum, and
that the ionized gas is located outside the QSO host galaxy.

The width of the emission lines implies that the broad--line region is
completely obscured from our line of sight, and we could be seeing the
ionized gas in the intermediate density narrow line region (NLR,
$n_H\sim 10^4-10^6$ cm$^{-3}$).  Typical average values of the line
ratios for radio galaxies at $2<z<3$ are $\langle
{Ly\alpha}/HeII\rangle=9.80\pm5.69$ and $\langle
CIV/HeII\rangle=1.50\pm 0.56$
\citep{mccarthy1993,humphrey2008}. \citet{matzuoka2009} provides the
average \civ/\heii\ ratio in bins of \heii\ luminosity and
redshift. For $2<z<2.5$ and $41.5<log(L_{HeII})<42.5$ the average
\civ/\heii\ ratio of the NLR is $1.34^{+0.57}_{-0.4}$. In the
composite spectrum calculated by \citep{humphrey2008} the \civ/\heii\
ratio is 1.75. Therefore the observed line ratios in the spectrum of
Ly1 (\civ/\heii$<0.8$, \lya/\heii$=2.22\pm 0.98$) is inconsistent with
the typical high-redshift NLRs at the 90\% level. We also note that
the typical line width of the \heii\ emission line in HzRGs is of the
order of $\ge 500$ \kms\ \citep{humphrey2008,villar-martin1999}.

Shocks could provide an alternative mechanism in cases where large
scale outflows and/or radio jets interact with the host galaxy
ISM. However, the low \civ/\heii\ ratio can only be produced by shock
models when the shock speed exceeds $\sim 400$ \kms, which is
inconsistent with the low velocity dispersion measured in the \lya\
nebula \citep{allen2008}.

\citet{matzuoka2009} recently presented a correlation between the
\heii\ luminosity (used as a proxy for the AGN luminosity) and the
metallicity of the NRL in high redshift radio galaxies (HzRGs). If
produced within the NLR, the observed $L_{HeII}=7.9\times 10^{41}$ erg
sec$^{-1}$ would imply a gas metallicity of $\sim 0.8 Z_{\odot}$, with
a large scatter. In their calculations they assume a gas density
$n_H=10^4$ cm$^{-3}$, typical of the intermediate density nuclear NLR
\citep{nagao2001,nagao2002}. They also find that the measured
metallicity of the NLR would not change significantly if the density
was two order of magnitudes lower ($n_{H}=10^2$ cm$^{-3}$). We can use
this metallicity to compare the observed \civ/\heii\ limit with the
the photoionization models calculated by \citep{villar-martin2007} for
the NLR of HzRGs.  For metallicity $Z=0.8Z_{\odot}$ and gas density
$n_H=10^4$ cm$^{-3}$, the \citet{villar-martin2007} models predict
that \civ/\heii\ ratio should be equal to 1.4 and 2.4, for a
ionization parameter\footnote{$U=Q/4\pi r^2nc$, where $Q$ is the
  photo-ionizing luminosity, $r$ is the distance between the gas and
  the ionizing source, $n_H$ is the total Hydrogen gas density, and
  $c$ is the speed of light.}  $U$ of 0.7 and 0.05,
respectively. These ratios are inconsistent with the observed limit.
Lower density models ($n_H \le 10^2$ cm$^{-3}$, typical of low density
extended emission line region) predict lower luminosity of the
collisionally excited \civ\ line. However, densities as low as $n_H=
10$ cm$^{-3}$ result in \civ/\heii$\ge 0.8-1.3$, for $U=0.7-0.05$
\citet{villar-martin2007}.

If the \heii\ emitting gas is located outside Ly1 it is most likely
associated with the extended \lya\ nebula. The \heii\ luminosity
implies a rate of $He^+$ ionizing
photons\footnote{$L_{He~II}=5.67\times 10^{-12} Q(He^+)$,
  \citep{schaerer2003}}, $Q(He^{+})$, of $\sim 1.5\times 10^{53}$
photons sec$^{-1}$. As recently suggested by \citet{geach2009} a
fraction of the QSO ionizing continuum could escape the nucleus
through an un-obscured direction, and could be enough to power the
\lya\ nebula. To produce the \heii\ line, the un-reddened continuum
should have a rest-frame luminosity at 1350\AA\ of $L_{1350\AA}\sim
4.5\times 10^{40}$ erg sec$^{-1}$ \AA$^{-1}$, within a factor of two
of the measured luminosity of Ly1 at 1350\AA. Such a luminosity,
however, would still be a factor of at least three too low to account
for the total \lya\ luminosity in the nebula. In fact, the \lya\
luminosity has to be considered as a lower limit, due to the effect of
\lya\ absorption \citep[e.g, ][]{prescott2009}.

Recently \citet{prescott2009} presented the discovery of a giant
$z=1.63$ \lya$+$\heii\ emission line nebula. They are able to
reproduce the observed ratio of \civ/\heii$=0.22$ with extremely metal
poor gas ($Z\le 10^{-2}-10^{-3} Z_{\odot}$) ionized by a hard
continuum, under certain assumptions for the unconstrained ionization
parameter. In our system, however, it is hard to reconcile such a low
metallicity gas with the fact that both Ly1 and Ly2 contain
significant amount of dust.

From the above discussion we conclude that the gas producing the
observed emission line spectrum is most likely not photoionized by the
hard continuum of the active nucleus. However, the various parameters
entering the photoionization models are largely unconstrained, and
more observations are needed to conclusively rule out the AGN
interpretation.

\subsubsection{Evidence against metal poor stars}

Can the radiation produced by stars be responsible for the observed
\lya\ and \heii\ emission?  \citet{schaerer2003} and
\citet{panagia2003} show that significant nebular \heii\ emission in a
cloud photoionized by stars can only be observed in the presence of
stars with very low metallicity ($\log(Z/Z_{\odot})<-5.3$) and/or
Population~III objects, due to their high effective temperature, and
high masses \citep{tumlinson2000,abel1998}. Normal stellar populations
emit hardly any flux below $\lambda =228 $\AA, and so cannot explain
the \heii\ emission. Does an exotic population of metal poor stars
reside in the starburst galaxy Ly2?

Although it has been suggested that population~III stars might be
present at redshifts as low as 3 \citep[see, e.g., ][]{jimenez2004},
star formation in metal free gas pockets is expected to be very rare
and improbable at $z\sim 2.5$ \citep{scannapieco2003}. More
importantly, the large amount of energy emitted by both Ly1 and Ly2 at
IR wavelengths, together with the detection of PAH emission, imply
very large masses of dust.  Therefore, a substantial amount of star
formation must have already happened in order to produce it. For the
population~III scenario to work, the population~III starburst must be
happening in a pocket of gas unpolluted by metals, while still being
attenuated by a screen of dust created in a previous episode of star
formation. This combination seems too contrived to be plausible,
especially in light of numerical simulations showing that gas mixing
is very efficient within a galaxy after the first episodes of star
formation \citep{tornatore2007}.  It is further instructive to examine
how much star formation would be required in such a pocket of
population~III stars.

To first approximation, the nebular emission lines are proportional to
the ionizing photon flux escaping the galaxy and (under some
assumptions about the IMF, metallicity, and age of the burst) to the
SFR of the galaxy \citep{schaerer2003}. The number of He$^+$ ionizing
photons, Q(He$^+$), is a strong function of the metallicity of the
stellar population, so we consider here only the extreme case of a
pure population~III starburst. For an escape fraction of 10\%
\citep[$f_{esc}=0.1$, ][]{iwata2008}, and neglecting the effect of
dust for the present, we find that the minimum star formation rate
required to power the \lya\ nebula is of $\sim 50 M_{\odot}$
yr$^{-1}$.  Since we neglected the dust absorption, this estimate
represents a lower limit to the true expected SFR in population~III
stars.  Given the arguments given above, population~III stars are
unrealistic, and we rule out the presence of metal poor stars as a
source of He$^+$ ionizing flux.

Finally, large-scale outflows driven by a strong starburst in a
forming galaxy were proposed as another viable mechanism to explain
the \lya\ nebulae \citep{taniguchi2000}. Recent high resolution
hydrodynamic simulations show that multiple SN explosions in
starbursts of a few hundred of $M_{\odot}$ yr$^{-1}$ could result in
large scale intense \lya\ emission, comparable with the observed one
in \lya\ nebulae \citep{mori2004}. However, in these models the
configuration of the system is rather symmetrical, with the star
forming region located at the center of the ionized gas. From
Figure~\ref{fig:images} one can see that the starburst galaxy (Ly2) is
located at the edge of the main \lya\ emitting region.

\subsubsection{Cooling radiation?}

Another possibility, consistent with the observations, is that the
extended \lya\ emission is coming from gas losing gravitational energy
while falling on a galaxy-size dark matter halo. \citet{fardal2001}
and \citet{keres2005} cosmological simulations show that the majority
of such infalling gas never reaches the virial temperatures expected
for the dark matter halo mass, $\sim 10^6$~K (which would emit
primarily in the X-ray), but rather cools to lower temperatures of a
few $10^5$~K. At these temperatures, both \lya\ and \heii\ emission
are expected. Although the \lya\ is expected to be the strongest line,
problems related to the radiative transfer of the \lya\ photons, the
details of the velocity field of the IGM, and the self shielding of
the gas at high densities, make the \lya\ predictions very uncertain
in these models. A distinctive feature of the cooling radiation is
that He lines should be observed, with the \heii\ emission coming
mostly from collisionally excited He$^{+}$ \citep{yang2006}. We note
however that these models assume primordial compositions for the
gas. The presence of metals has an impact on the gas cooling function
possibly affecting the predicted intensity of the \heii\ emission
line.  In the \citet{yang2006} models the \heii\ line is expected to
have velocity widths not larger than $\sigma\sim 300$\kms, and a
distribution clumpier than that of the \lya\ emission. For these
reasons, even if the \heii\ emission is expected to be several factors
fainter than the \lya, it can still be detected with observations
similar to those presented here.

Using the \lya\ luminosity, we can estimate the mass deposition rate
($\dot{M}$) due to the cooling gas. The \lya\ luminosity of $\sim
6\times 10^{43}$ erg s$^{-1}$ implies $\sim 10^{61}$ recombination per
year, or $\dot{M}\sim 10^4/n\, M_{\odot}$ yr$^{-1}$, where $n$ is the
number of recombinations per H atom. By considering $n$ in the range
between 10--100 \citep{heckman1989} we find a mass deposition rate in
the range between $\sim 10^2 - 10^3 M_{\odot}$ yr$^{-1}$.  The cooling
material would then be enough to fuel both the QSO activity and the
high star formation rate observed in Ly2.

\section{Conclusions}
\label{sec:conclusions}
We have presented a multi-wavelength study of an extended \lya\ nebula
identified in an over-dense filamentary structure at redshift $z=2.38$
\citep{palunas2004}. Two sources (Ly1 and Ly2) of continuum are found
to be physically associated with the extended \lya\ emission.  Both
are powerful sources detected at 24$\mu$m, but not detected at either
$850\mu$m or radio wavelengths, nor in the X-ray.  The power-law IR
SED of Ly1 is consistent with that of a heavily reddened QSO, while
Ly2 is a starburst galaxy, with properties similar to other powerful
star-forming galaxies at $z\sim 3$. The UV rest-frame spectra of these
two objects confirm that the two galaxies belong to the
$z=2.38$\ structure.  They have only a small difference in velocity (a
few hundred \kms), so it is possible that we are witnessing an early
stage of a merger between two massive galaxies.

The rest-frame UV spectrum observed at the position of Ly1 shows
emission from \heii\ at 1640~\AA, while we find no associated \civ\ or
N~{\small V} lines. Although it is possible that the observed emission
line spectrum is generated in gas ionized by the continuum of the QSO,
the the observed \civ/\heii\ limit suggests that the \lya\ and the
\heii\ radiation are the result of a different mechanism. The \heii\
line width argues against a WR stellar origin of the emission line.
Extremely metal poor stars can be excluded as a source of ionization
because of the presence of metal absorption lines in the UV spectrum
of Ly2, and by the presence of dust inferred from the detection of
both sources at 24$\mu$m and the strong PAH lines in Ly2.

Another possibility that would be able to explain the observations is
that the \heii\ and the \lya\ emission are produced by collisions in
gas that is being accreted on the dark matter halo of the two
galaxies, Ly1 and Ly2. Future rest-frame optical spectroscopy
will help in the interpretation of the \lya\ cloud. If due to cold
accretion, then the \lya\ is caused by collisional excitation rather
than photoionization. In this case, the \ha\ line should be only about
2\% of the \lya\ line \citep{fardal2001}. Furthermore, the ratio
between strong optical ([OIII], \hb, \ha, and [N~{\small II}])
emission lines could be used to investigate the source of ionization
in the nebula.

Given the complexity of the environment associated with our \lya\
nebula it is possible that the various mechanisms of excitation are at
work simultaneously. This possibility has been discussed recently by
\citet{dijkstra2009}, who point out that the association of cold
accretion with powerful starburst and/or AGN has to be expected, since
ultimately these sources are triggered and fed by the gas infall.

Although theoretically the cold mode of accretion is expected to be
the dominant mechanism for growth in massive galaxies at $z\sim 2$
\citep{dekel2008}, we still lack a direct observational confirmation
of its existence. A common prediction among the different models of
cold accretion is the unique filamentary morphology expected for the
emitting gas. The typical filaments are predicted to have a surface
brightness of about $10^{-19.5}$ erg s$^{-1}$ cm$^{-2}$ arcsec$^{-2}$
in both \lya\ and \heii, and the cooling luminosity is expected to
peak at redshifts between $\sim 1$ and 2. At these redshifts, the
\lya\ line is barely accessible from the ground and current space
based facilities are not sensitive enough to reach the low surface
brightness levels in reasonable exposure times. (Reaching a surface
brightness of $10^{-19}$ erg s$^{-1}$ cm$^{-2}$ arcsec$^{-2}$ in the
observed \lya\ line at $z=1.8$ with the Hubble Space Telescope Wide
Field Camera 3 UVIS channel would require $\sim 10^7$s). The proposed
Advanced Technology Large-Aperture Space Telescope (ATLAS) could
potentially detect cooling radiation in the redshift range (between
$z=$1 and 2) where cold flows are expected to be the dominant
mechanism for gas accretion in galaxies.

\acknowledgments We thank the referee for useful suggestions that
improved the discussion of the results.  We thank A. Pope for
providing the templates in electronic form. C.S. thanks M. Hayes for
useful discussions.

\bibliographystyle{apj} 
\bibliography{scarlatabib}

\clearpage
\begin{table}
\begin{center}
\caption{Optical image quality and flux limits.\label{tab:data}}
\begin{tabular}{lccl}
  \tableline\tableline
  Band  & FWHM\tablenotemark{(a)} & $5\sigma$ &Reference\\
  & [$''$] &  & \\
  \tableline
  \tableline
 % Xray & XXX & $5\times10^{-15}$\tablenote{In \ecs.}&Williger et al. (XXXX)\\
  $U$ &1.1 &26.0\tablenotemark{(b)}&\citet[][in preparation]{scarlata2009}\\
  $B$ &1.4 &26.2&Palunas et al. (2004)\\
  $R$ &1.3 &24.0&\citet[][in preparation]{scarlata2009}\\
  $J$ &0.9 &22.9&\citet[][in preparation]{scarlata2009} \\
  $H$ &0.9 &22.7& \\
  3.6$\mu$m  &1.7 &24.3&\citet[][in preparation]{colbert2009}\\
  4.5$\mu$m  &1.7 &23.5&\\
  5.8$\mu$m  &1.9 &21.4&\\
  8.0$\mu$m  &1.9 &21.2&\\
  24$\mu$m  & 5.4 &19.4&\citet{colbert2006} \\
  870$\mu$m & &8\tablenotemark{(c)} &\citet{beelen2008} \\

  \tableline
  \tablenotetext{(a)}{In the optical and near-IR images the magnitude
    limits were computed within an aperture of $1.5\times$FWHM
    diameter.}
  \tablenotetext{(b)}{In AB magnitudes.}
  \tablenotetext{(c)}{In mJy.}
\end{tabular}
%% Any table notes must follow the \end{tabular} command.
\end{center}
\end{table}

\begin{center}
\begin{deluxetable}{lcccccccccc}
\tablecolumns{11} 
\tablewidth{0pc} 
\tablecaption{Broad band photometry (AB magnitudes).\label{tab:photometry}} 
\tabletypesize{\scriptsize}
\tablehead{ 
\colhead{Objects}    &  
\colhead{$U$}   & 
\colhead{$B$}    &
\colhead{$R$}    &  
\colhead{$J$}   & 
\colhead{$H$}    &
\colhead{3.6$\mu$m}    &  
\colhead{4.5$\mu$m}   & 
\colhead{5.8$\mu$m}    &
\colhead{8.0$\mu$m}    &  
\colhead{24$\mu$m}   
}

\startdata
Ly1 &  25.9$\pm$0.3& 24.9$\pm$0.1&$>24.5$&  21.2$\pm$  0.1&20.8$\pm$ 0.1&  19.8$\pm$ 0.5& 19.5$\pm$0.4& 18.9$\pm$0.3& 18.2$\pm$0.3& 17.0$\pm$0.3 \\
Ly2 &  24.5$\pm$0.1& 23.8$\pm$0.1& 23.4$\pm$0.2& 22.6$\pm$  0.3&21.5$\pm$ 0.1&  20.2$\pm$ 0.6& 20.1$\pm$0.4& 19.8$\pm$0.5& 19.8$\pm$0.5& 16.9$\pm$0.3\\
\enddata

\end{deluxetable}
\end{center}

\begin{table}
\begin{center}
\caption{Emission line properties.\label{tab:spectroscopy}}
\begin{tabular}{lccc}
  \tableline\tableline
  & Flux & $EW$ & $FWHM$\tablenotemark{(a)}\\
  &erg cm$^{-2}$ s$^{-1}$ &\AA\ & km s$^{-1}$\\
  \tableline
  Ly1 & $z=2.373$& & \\
  \tableline
  \lya\ &5.0 $\pm 0.7\times 10^{-17}$&131.2$\pm 25.9$&$< 366$\\
  \heii\ &1.8 $\pm 0.7\times 10^{-17}$&14.2$\pm 4.1$& $< 400$\\
  \civ\ & $<1.5\times 10^{-17}$\tablenotemark{(b)}& -- & -- \\
  \tableline\tableline
  Ly2 & $z=2.367$& &\\
  \tableline
  \tableline
  \lya\ &7.3 $\pm 0.7\times 10^{-17}$&67.3$\pm 19.7$&202\\
  \tableline
  \tablenotetext{(a)}{The FWHM is not corrected for instrumental resolution.}
  \tablenotetext{(b)}{2$\sigma$ limit, see text for details.}
\end{tabular}

\end{center}
\end{table}

\end{document}